\documentclass[aps,prl,twocolumn,superscriptaddress,showpacs,amsfonts]{revtex4-1}
\usepackage{epsfig}
\usepackage{graphicx}
\usepackage{amsmath}
\usepackage{amssymb}
\usepackage{color}
\usepackage{bm}

\newcommand{\beq}{\begin{equation}}
\newcommand{\eeq}{\end{equation}}
\newcommand{\beqarray}{\begin{eqnarray}}
\newcommand{\eeqarray}{\end{eqnarray}}

\begin{document}

\title{Driving Topological Phases by Spatially Inhomogeneous Pairing Centers}

\author{Wojciech Brzezicki}
\affiliation{CNR-SPIN, IT-84084 Fisciano (SA), Italy}
\affiliation{Dipartimento di Fisica ``E. R. Caianiello\textquotedblright{},
 Universit\'a degli Studi di Salerno, IT-84084 Fisciano (SA), Italy}

\author{Andrzej M. Ole\'s}
\affiliation{Max Planck Institute for Solid State Research,
             Heisenbergstrasse 1, D-70569 Stuttgart, Germany}
\affiliation{\mbox{Marian Smoluchowski Institute of Physics, Jagiellonian University,
             prof. S. {\L}ojasiewicza 11, PL-30348 Krak\'ow, Poland}}

\author{Mario Cuoco}
\affiliation{CNR-SPIN, IT-84084 Fisciano (SA), Italy}
\affiliation{Dipartimento di Fisica ``E. R. Caianiello\textquotedblright{},
 Universit\'a degli Studi di Salerno, IT-84084 Fisciano (SA), Italy}

\date{\today}

\begin{abstract}
We investigate the effect of periodic and disordered distributions of
pairing centers in a one-dimensional itinerant system to obtain the
microscopic conditions required to achieve an end Majorana mode and
the topological phase diagram. Remarkably, the topological invariant
can be generally expressed in terms of the physical parameters for any
pairing center configuration. Such a fundamental relation allows us to
unveil hidden local symmetries and to identify
trajectories in the parameter space that preserve the non-trivial
topological character of the ground state. We identify the phase
diagram with topologically non-trivial domains where Majorana
modes are completely unaffected by the spatial distribution of the
pairing centers. These results are general and apply to several systems
where inhomogeneous perturbations generate stable Majorana modes.
\end{abstract}
\maketitle

\textit{Introduction.}--- A topological phase is generally marked by
quantized macroscopic observables which are insensitive
to modifications of the local environment
\cite{Hasan2010,Qi2011,Chi16} as for the well-known example of the
quantum Hall effect \cite{Klitzing1980,Thouless1982}.
In many cases, however, there is a limitation
on the allowed perturbations, leading to so-called symmetry protected
topological phases \cite{Kitaev2009,Schnyder2009,Wen2014}.
Strong topological phases are due to global symmetries, such as
time-reversal, particle-hole, or chirality, while translation or point
group symmetries of the lattice
lead commonly
to weak and crystalline
topological states \cite{Fu2007,Moore2007,Roy2009,Fu2011,Hsieh2012}.
Weak or strong character refers to what extent the protecting symmetry
can be achieved in a realistic configuration. Indeed, while time reversal
symmetry can be controlled by avoiding magnetic impurities,
for weak topological phases the translation symmetry can be
broken by impurities in a crystal.
However, even when disorder breaks the lattice symmetry, a weak
topological phase can still be robust if the system remains symmetric
on average \cite{Nomura2008,Mong2012,Fu2012,Kobayashi2013,Fulga2014,Baireuther2014,Sbierski2014,Diez2015,Morimoto2015}.

Though harmful for some types of topological protections, inhomogeneous
perturbations have recently provided new perspectives as a rich
intrinsic source of topological phases or topological transitions.
Some relevant examples of this type are topological Anderson
insulators \cite{Li2009,Groth2009} and disorder driven topological
superconductivity \cite{Adagideli2014,Qin2015}. Impurities with a
periodic pattern placed on superconductors or insulators can generally
lead to robust zero energy crossing for increasing impurity strength
\cite{Kimme2016}, enabling, among others,
topological phases with very large Chern numbers \cite{Rontynen2015}.
Thus the use of magnetic impurities represents
a consolidated route to manipulate or induce topological phases.
For instance, magnetically active dopants on the top of a topological
insulator \cite{Liu2009,Biswas2010,Chen2010,Schlenk2013} are employed
to break time reversal symmetry and give rise to a topological
magnetoelectric effect \cite{Qi2008}. Even more striking is the example
of a one-dimensional (1D) topological superconductor (TSC)
obtained through the deposition of a chain of magnetic ad-atoms on the
surface of a superconductor when they order magnetically, even if the
superconductor is topologically trivial \cite{Choy2011,Nadj-Perge2011,Nakosai2013,Klinovaja2013,Braunecker2013,Vazifeh2013,
Pientka2013,Poyhonen2014,Pientka2014,Li2016,Heimes2014,Heimes2015,Nadj-Perge2014,Pawlak}.

Remarkably, every TSC can be linked to a $p$-wave superconductor (PWS)
in a suitable limit \cite{Altland,Kitaev2009,Tewari2012,Ryu}, as for the
paradigmatic case of electrons forming Cooper pairs in the symmetric
spin-triplet and orbitally-antisymmetric configuration. Such a
connection motivated the proposal of artificial TSCs
\cite{FuKane2008,AliceaReview,BeenakkerReview,FlensbergReview,KotetesClassi}
and the subsequent observation of Majorana modes in hybrid
superconducting (SC) devices
\cite{Mourik,Deng,Furdyna,Heiblum,Finck,Churchill,Franke,Pawlak,Albrecht}.
In this context, the issue of disorder in the employed effectively
spinless PWSs is of great relevance especially in view of any realistic
implementation in devices for topological quantum computing.
After the pioneering work of Ref. \cite{Motrunich2001},
this problem has been largely explored in the literature, mainly
focusing on spatially varying charge potential for periodic \cite{DeGottardi2011,Niu2012,Sau2012,DeGottardi2013}, quasiperiodic \cite{DeGottardi2013,Tezuka2012,Lang2012}, or disordered
\cite{Brouwer2011,Lobos2012,Cook2012,Pedrocchi2012} patterns,
indicating that a sufficient strength of
SC correlations
is basically required to drive the system into a topological phase.

Apart from local charge density disorder, there is another fundamental
route to explore the intricate relation between inhomogeneous
perturbations and topological behavior in such a class of systems
where Majorana modes may occur at the edge. Indeed, impurities can be
introduced into a system of itinerant electrons as effective pairing
centers (PCs), both by artificial or intrinsic means, thus focusing
attention on the role of the profile of the pairing amplitude rather
than of the charge density.
Such a problem has a broader physical context if one considers
the formation in a metallic host of local electronic states that
prefers to be either empty or with two paired electrons.
Indeed, due to phononic \cite{She2013,Fransson2013} or excitonic
\cite{Varma1988} origin,
PCs can lead to both pairing mechanisms and superconductivity
\cite{Varma1988,Micnas1990,Bar-Yam1991,Dzero2005} in various materials
as well as drive superfluid-to-insulator transitions \cite{Cuoco2004}.
Furthermore, impurities on the surface of topological insulators or
Dirac materials \cite{She2013,Fransson2013} have also been suggested
as generators of local PCs. Finally, the same physics may arise in
spin-orbital coupled quantum systems with
spatially inhomogeneous anisotropic exchange in the presence of
impurities with different valence \cite{Brz15,Brz16}.

In this Rapid Communication, we investigate the effect of periodic
and disordered (within a large unit cell) distributions of effective
$p$-wave PCs in a 1D electron system to obtain the microscopic
conditions required for the existence of an end Majorana mode.
Remarkably, the topological invariant can be generally expressed in
terms of the physical parameters for any
distribution of the PCs.
A striking consequence of the emerging
symmetries in the parameter space is the finding of a physical
regime where Majorana modes can be completely unaffected by the
overall spatial distribution of the PCs.

{\it The model.}--- We investigate a ring described by a 1D
tight-binding model of spinless electrons with an inhomogeneous
distribution of PCs. The Hamiltonian is a modification of the one
originally introduced by Kitaev \cite{Kitaev2001}, and includes
inhomogeneities generated by diluted PCs distributed in the unit cell
of length $L$.
Using system periodicity which introduces momentum $k$ we get,
\begin{equation}
{\cal H} =\!\sum_{p=1\atop k}^{L}\!
\left\{t_{p}^{} c_{k,p}^{\dagger}c_{k,p+1}^{}
\!+\!\Delta_p^{} c_{k,p}^{\dagger}c_{-k,p+1}^{\dagger}\!+\!\textrm{H.c.}
\!+\!\mu_{p}^{} c_{k,p}^{\dagger}c_{k,p}^{}\right\}\!,
\end{equation}
with boundary conditions $c_{L+1,k}\equiv e^{ik}c_{1,k}$ and
$\{t_{p},\Delta_{p}\}$ being the nearest neighbor (NN) hopping and
on-bond pairing amplitudes, respectively. We assume that there are
$N$ impurities in the unit cell labeled by $i$ at generic
(but non-neighboring) positions $\{p_i\}$ along the chain
such that $\Delta_{p}\neq 0$ only at the bonds around them,
i.e., $\Delta_{p_i-1}=\Delta_{p_i}\equiv\Delta_i$.
Hopping and chemical potential for the host subsystem take
uniform values, $t_p\equiv t_0$ and $\mu_p\equiv \mu_0$
(which can be transformed to an alternating $\mu_p=(-1)^p\mu_0$,
see the Supplemental Material \cite{Sup}),
while at the host-impurity bonds these parameters
are given by arbitrary amplitudes,
$t_{p_i-1}=t_{p_i}\equiv t_i$ and $\mu_{p_i}=\mu_i$, see Fig. 1.
The Hamiltonian belongs to the BDI class of the Altland-Zirnbauer
classification \cite{Altland} and it can have a non-trivial topological
phase characterized by a $\mathbb{Z}$ winding number
--- finite $\mathbb{Z}$ implies end Majorana modes in an open chain
as explicitly demonstrated in the Supplemental Material \cite{Sup}
for a representative case.

For real values of parameters, due to the chiral symmetry, $\cal H$ Eq.(1)
can be casted into a purely block-off diagonal form with antidiagonal
blocks given by matrices ${\bf u}_k$ and ${\bf u}_k^{\dagger}$. Hence,
as long as the eigenvalues of ${\bf u}_{k}$ are gapped, its determinant
$\det{\bf u}_{k}$ can be used to get the winding number $\mathbb{Z}$ by
evaluating its trajectory in the complex plane. We observe that the
phase of the determinant is a topological invariant because it is not
related to any symmetry breaking and it can only change when it goes
through zero. In general, by Laplace transformation we have that
$\det{\bf u}_k=\alpha({\cal A}+{\cal B}\cos k+i\,{\cal C}\sin k)$, with
$\alpha$, $\cal A$, $\cal B$, and $\cal C$ being real coefficients that
are independent of~$k$. Then, a general condition for a non-trivial
$\mathbb{Z}$
is provided by having both:
(i)~$|{\cal A}|<|{\cal B}|$ and
(ii)~${\cal C}\neq 0$,
giving
$\mathbb{Z}=\pm 1$ for positive/negative ${\cal C}$.

\begin{figure}[t!]
\begin{center}
\includegraphics[width=1.0\columnwidth]{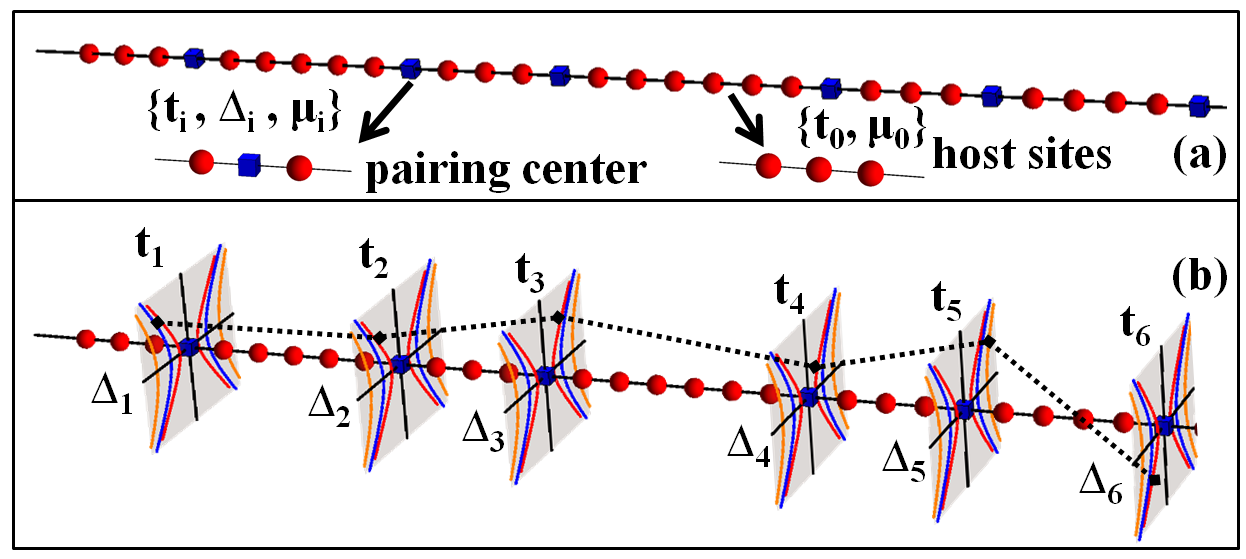}
\end{center}
\vskip -.3cm
\caption{
(a) Schematic representation of
a large unit cell of length $L=30$ sites with $N=6$ impurities within
the 1D chain with spatially inhomogeneous distributed
PCs (blue squares) which separate host sites (red circles) and
reduce the translational invariance.
(b)~A~representative trajectory (dashed line) in the parameter space
connecting different impurities in~(a).
Local variation along the hyperbolic contours in the shadow plane at
the impurity site can lead to topologically equivalent non-trivial
trajectories.
}
\label{fig:1}
\end{figure}

{\it Symmetry features of the topological invariant.}---
The central result obtained in this paper is the general analytical
expression for the topological invariant of an inhomogeneous quantum
system. This
outcome allows us to unveil the emergent symmetries in the
parameter space and to predict the key physical regimes for the
occurrence of Majorana end modes. By means of an inductive construction
and a recursive relation in terms of the number of impurities within
the unit cell, the coefficient $\cal A$ is
\begin{eqnarray}
{\cal A} & \!= &\cos\left(L\eta_{0}\right)\!
+\sum_{i=1}^{N}y_{i}\sin\!\left(L\eta_{0}\right)\nonumber \\
& \!+ & 2^{1}\sum_{i<j}y_{i}y_{j}\sin\!\left(d_{ij}\eta_{0}\right)
\sin\!\left(d_{ji}\eta_{0}\right)               \nonumber \\
& \!+ & 2^{2}\!\sum_{i<j<k}\!\! y_{i}y_{j}y_{k}\sin\!
\left(d_{ij}\eta_{0}\right)\sin\!\left(d_{jk}\eta_{0}\right)
\sin\!\left(d_{ki}\eta_{0}\right)+\dots         \nonumber \\
& \!+ & 2^{N-1}y_{1}y_{2}\cdots y_{N}\!
\sin\!\left(d_{12}\eta_{0}\right)\!\sin\!\left(d_{23}\eta_{0}\right)
\cdots\sin\!\left(d_{N1}\eta_{0}\right),        \nonumber \\
\label{eq:master}
\end{eqnarray}
where $d_{ij}$ are the {\it ordered} distances between impurities
which cover the unit cell for each term of this expansion,
i.e., $d_{ij}\equiv p_j-p_i$ for $j>i$ and $d_{ji}\equiv L-p_j+p_i$
for the second line, \textit{etcetera}, see the Supplemental Material
\cite{Sup}. The above equation can be written in a compact form,
${\cal A}=\cos(L\eta_{0})+{\rm Tr}\{({\bf 1}
-2{\bf M}_{1}{\bf Y})^{-1}{\bf M}_{2}^{T}{\bf Y}\}$,
in terms of two triangular matrices of impurities' distances
${\bf M}_{1}$ and ${\bf M}_{2}$, and a diagonal matrix ${\bf Y}$.
These matrices have the following non-vanishing entries:
$({\bf M}_{1})_{ij}=\sin(\eta_{0}d_{ij})$,
$({\bf M}_{2})_{ij}=\sin(\eta_{0}(L-d_{ij}))$ for $j\geq i$
and ${\bf Y}_{ii}=y_i$.
Here $\eta_0=\arccos(\mu_0/(2 t_0))$ is the effective inverse Fermi
length of the host, and $y_{i}$ are dimensionless variables related to
the parameters at the impurity and in the host via
\begin{equation}
y_{i}=\frac{t_{0}(J_{i}^{-1}-J_{0}^{-1})}{\sqrt{4-\mu_{0}^{2}/t_{0}^{2}}},
\end{equation}
with $J_{i}=(t_{i}^{2}-\Delta_{i}^{2})/\mu_i$ being a renormalized
energy scale that includes effectively the particle- and hole-like
transfer processes between the impurities and the host.
Similarly, for the host one has $J_{0}=t_{0}^{2}/\mu_{0}$. Concerning
${\cal B}$ and ${\cal C}$, it is convenient to parametrize the set
$\{t_i,\Delta_i,\mu_i\}$ at each host-impurity bond using hyperbolic
coordinates as $t_{i}=r_{i}\cosh\phi_{i}$, $\Delta_i=r_i\sinh\phi_i$,
and $\mu_i=r_i^2 J_i^{-1}$, because they explicitly manifest their
unique dependence on the hyperbolic angles as:
\begin{equation}
{\cal B}=\cosh\left(2\sum_{i=1}^{N}\phi_{i}\right),\quad
{\cal C}=\sinh\left(2\sum_{i=1}^{N}\phi_{i}\right).
\label{eq:BC}
\end{equation}

Hence, the topological invariant has a highly symmetric structure
in the parameter space. The coefficient ${\cal A}$ depends both on
impurity-host and host bonds whereas the coefficients ${\cal B}$ and
${\cal C}$ contain only information about the impurity-host bonds
through the sum of all hyperbolic angles. Note that the same properties
hold in a general case of $\{t_p,\Delta_p,\mu_p\}$ taking arbitrary
value on every bond/site, see the Supplemental Material \cite{Sup}.
Benefiting of
the analytical expression of the winding number we can easily unveil
hidden symmetries for the trajectories in the parameter space.

Firstly, a variation in the local angles $\{\phi_{i}\}$, akin to a
relativistic Lorentz rotation, will not affect the amplitude of
${\cal B}$ as far as the sum of all angles stays unchanged. In this
respect a given value of ${\cal B}$ corresponds to many equivalent
trajectories in the $\{t_i,\Delta_i,\mu_i\}$ multidimensional space,
see Fig. 1(b). In general it is always possible to turn the system
topological by changing one hyperbolic angle to satisfy the condition
$|{\cal A}|<|{\cal B}|$. This uncovers a novel non-local way to employ
inhomogeneous perturbations for either driving a topological transition
or for keeping unchanged the topological character of the ground state.
Second symmetry aspect
is related to the amplitude scaling of the impurity parameters.
Indeed, a local scaling of $\mu_{i}$ and $r_{i}^{2}$ by the same factor
will leave unchanged the amplitudes $\{y_i\}$. Moreover, either a
cyclic translation of all the impurities in the unit cell or having a
multiplied unit cell will not affect the coefficient ${\cal A}$.

\begin{figure}[t!]
\includegraphics[clip,width=1\columnwidth]{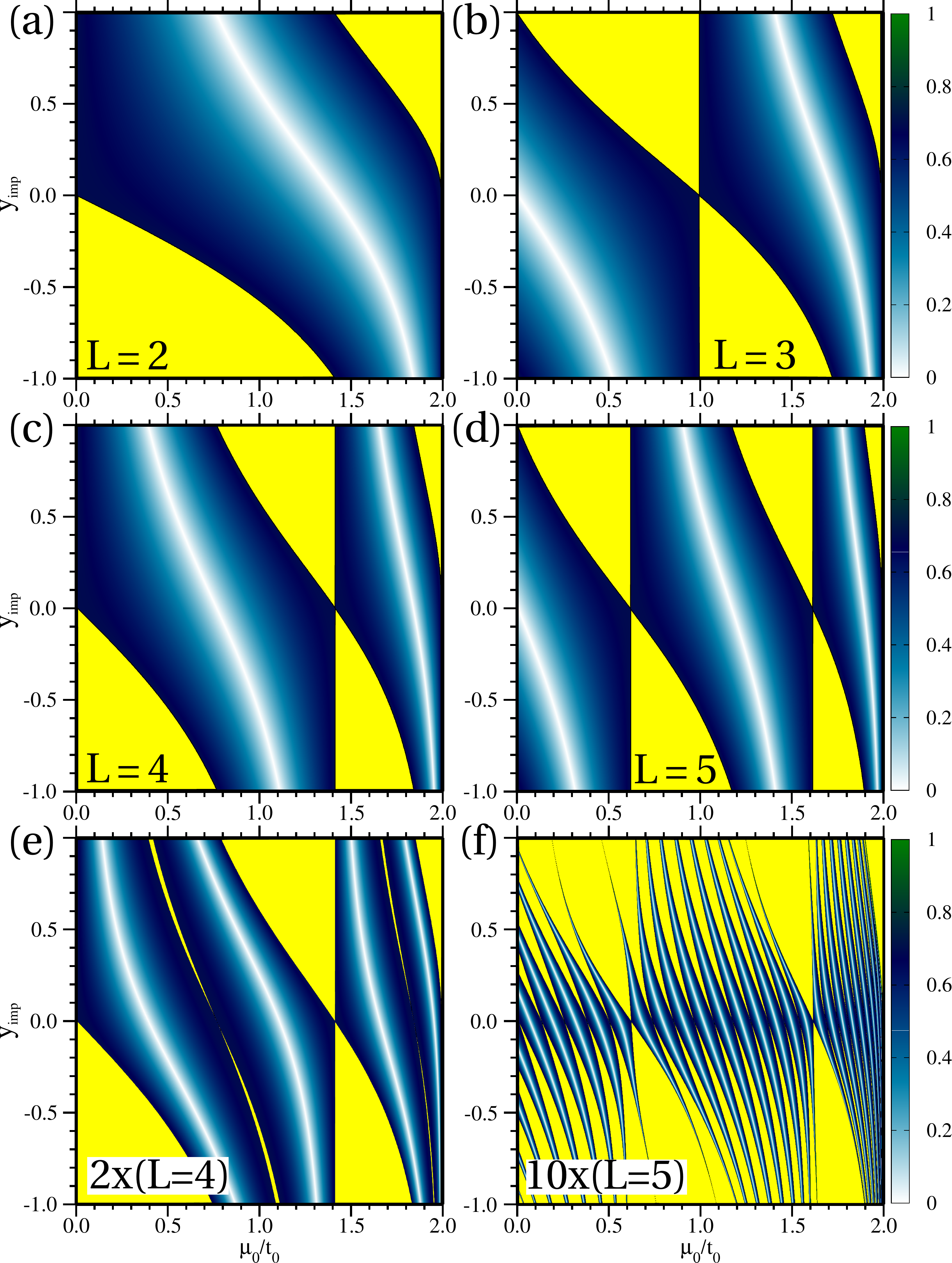}
\protect\caption{
Topological domains (blue regions) with: (a)-(d)~small
unit cells of the lengths $L=2,\dots,5$ with single impurity and
$y_i\equiv y_{\rm imp}$, (e) doubled unit cell of $L=4$ and
perturbed $y_i$, i.e., $y_1=y_{\rm imp}$ and $y_2=0.9\,y_{\rm imp}$
with impurities at regular position $p_{i}=4i$, and (f)~ten-fold unit
cell of $L=5$ and perturbed position of the last impurity by one bond,
i.e., $p_{i}=5i$ for $i=1,\dots,9$ and $p_{10}=51$ with
$y_i\equiv y_{\rm imp}$. Color map indicates values of $|{\cal A}|$
in the topological domains.
Yellow regions are for topologically trivial configurations.
\label{fig:reg2,3,4,5}}
\end{figure}

Finally, from the inspection of Eq. (\ref{eq:master}), we observe that
there exists a special point in the parameter space where all $\{y_i\}$
vanish, i.e., at $J_i\equiv J_0$ which is a resonance
condition between the host and impurities.
Then, a special critical point emerges in the phase diagram where the
interference between the impurities and the host makes
${\cal A}$ insensitive to the
actual spatial distribution of impurities.

{\it Periodic and disordered impurity pattern.}---
A topological domain can be obtained by imposing the condition
$|{\cal A}|<1$. By construction $|{\cal B}|\geq 1$ so for any choice
of hyperbolic angles $\{\phi_i\}$ such a region in the parameter
space will be topologically non-trivial.
For values $|{\cal A}|>1$ there can be regions with end Majorana mode,
nevertheless their boundaries would be exponentially unstable to small
changes in the parameters.
In Figs. \ref{fig:reg2,3,4,5}(a)-(d) we report the evolution of the
topological domains for a single impurity unit cell in terms of
the key physical parameters related to the impurity-host competition,
i.e., $y_i\equiv y_{\rm imp}$ and $\mu_0/t_0$.
The domains always extend around the line of impurity-host resonance
at $y_{\rm imp}=0$ for which $|{\cal A}|\le 1$.
We observe that the topological domains evolve around the nodal lines
of ${\cal A}$, and their number is $\lceil L/2\rceil$.
Finally, for all the cases the topological regions never extend
beyond $|\mu_0|=|2t_0|$ where the trigonometric functions in ${\cal A}$
become hyperbolic --- this corresponds to the
topological boundary of the pure Kitaev model~\cite{Kitaev2001}.

The question of having slightly different $y_i$'s at two PCs is
addressed in Fig. \ref{fig:reg2,3,4,5}(e) when we double the unit cell
of length $L=4$ [compare with Fig. \ref{fig:reg2,3,4,5}(c)] and set
$y_1=y_{\rm imp}$, $y_2=0.9\,y_{\rm imp}$.
We see that the topological domains split into halves with narrow
inclusions of the trivial phase and the number of nodal lines of
${\cal A}$ doubles. More splitting appear when we impose a small
perturbation of
impurity positions,
see Fig. \ref{fig:reg2,3,4,5}(f),
by taking a ten-fold unit cell of $L=5$
[compare with Fig. \ref{fig:reg2,3,4,5}(d)] and by moving one impurity
(out of $N=10$) by one site.

\begin{figure}[t!]
\includegraphics[clip,width=.95\columnwidth]{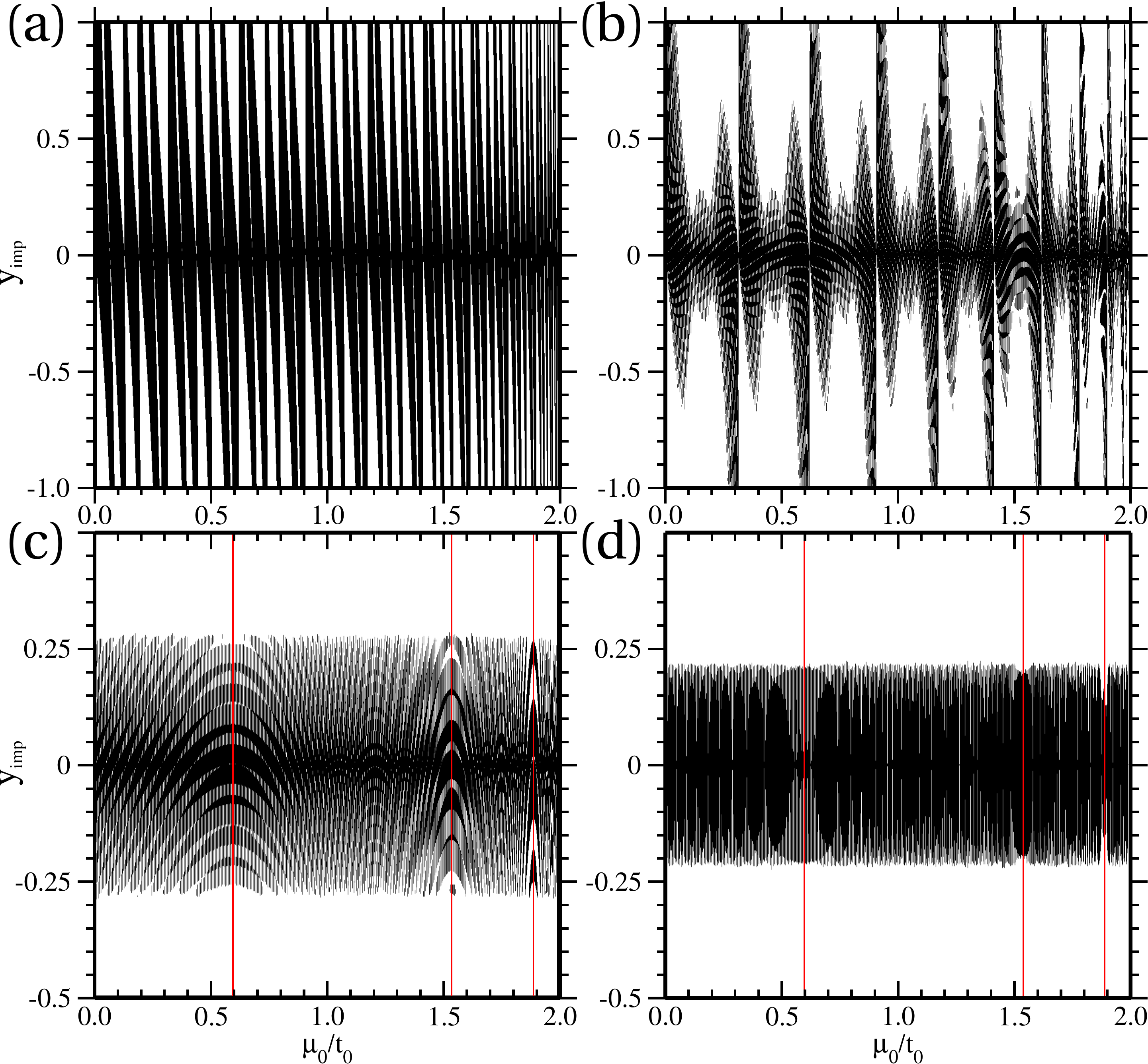}
\protect\caption{
Topological domains (black) for equivalent PCs,
$y_i\equiv y_{\rm imp}$, in:
(a) a dimerized system with alternating distance between NN
impurities $d_{1,2}\!=\!40$, $d_{2,1}\!=\!60$ and $L\!=\!100$,
(b) a system with the unit cell of $L\!=\!1000$ and $N\!=\!20$
impurities with NN distances taking random and
equiprobable value of $40$ or $60$, and
(c) the same system with random distribution of impurities and
$d_{ij}>1$. The random signs of $y_i$, i.e.,
$y_i\equiv\pm y_{\rm imp}$, are added to the case (c) in plot (d).
Red lines indicate $\mu_0/t_0 \simeq 0.6,1.54,1.9$,
for which a zoomed view of the interference fringes is reported in
the Supplemental Material \cite{Sup}.
\label{fig:dis_pos+Y}}
\end{figure}

Topological domains persist for dilute impurities in large unit cells
($N\ll L$) as seen
for a dimerized system with two distances between NN PCs,
$d_{12}=40$ and $d_{21}=60$ ($L=100$) and $y_i\equiv y_{\rm imp}$, see
Fig. \ref{fig:dis_pos+Y}(a).
Next we randomize the NN distances so that for every bond
$d_{i,i+1}=\{40,60\}$ with the same probability and implement the
constraint that the total number of short/long bonds is the same.
After averaging over $1000$ random configurations of unit cells
with $L=1000$ and $N=20$ we obtain a complex interference
pattern of many harmonics of $\eta_0$ which
destroys some of the vertical legs of initial topological region of
Fig. \ref{fig:dis_pos+Y}(a) and adds subtle parabolic modulation on
top of it, see Fig. \ref{fig:dis_pos+Y}(b).
Introducing complete randomness of the positions of PCs
modifies the domains further, see Fig. \ref{fig:dis_pos+Y}(c).
The long vertical legs are gone and the whole
topological regions extend not beyond
$|y_{\rm imp}|<0.25$. Interestingly the subtle parabolic features
remain with some characteristic points on the horizontal axis
(like $\mu_0/t_0=0.6,1.5,1.9$) around which the tips of many
vertically shifted parabolas seems to accumulate. We note that adding
even a significant random modulation of $y_i$ around $y_{\rm imp}$
does not change this result much as long as the signs of $y_i$
are fixed.

The parabolic features are suppressed and the domains gain
a mirror symmetry around $y_{\rm imp}=0$ when signs of $\{y_i\}$ are
random, see Fig. \ref{fig:dis_pos+Y}(d).
Quite surprisingly, the characteristic $\mu_0/t_0$ points still host
some distinct interference features. The overall width of the
topological window is further reduced but only slightly.
Note that lowering the concentration of PCs does not change much the
subtle features of topological regions of Figs.
\ref{fig:dis_pos+Y}(c) and \ref{fig:dis_pos+Y}(d) --- only the width
of the topological window is increased as the number of terms in
${\cal A}$ drops with $N$.

{\it Conclusions.}---
We provide a novel perspective on the way to get Majorana end modes in
1D itinerant systems by means of an inhomogeneous distribution of PCs.
The structure of the topological invariant uncovers
the reasons why domains in the parameter space can be so robust to the
local variations of impurity PCs. Such response to inhomogeneous
perturbations is a generalization of the Kitaev model within the
Altland-Zirnbauer classification and goes beyond the expectation from
density disorder.

An interesting aspect of our analysis points to the possibility of
having zero energy modes in the interior of the 1D diluted Kitaev chain.
Such occurrence can be captured by searching for a parameter's
configuration satisfying ${\cal C}=0$ and $|{\cal A}|\!<\!|{\cal B}|$. This
condition implies that the determinant $\det[{\bf u}_k]$ vanishes at a given $k$.
Hence, if the obtained zero energy states are not accidentally occurring at the boundary
they must be in the interior of the system.
Therefore, the combination of ${\cal C}=0$ and the
topological regions for which $|{\cal A}|\!<\!|{\cal B}|$ makes a guide
for the search of non-trivial states in the interior of the 1D diluted
Kitaev chain. Another path to engineer inhomogeneous topological phases
can be achieved by designing the system as a series of topologically
inequivalent domains. For instance, by selecting the microscopic
parameters for the impurity and the host such that the neighboring
domains have different winding numbers (e.g. they are topologically
inequivalent),
Majorana modes would occur at the domain boundary in the interior
of a quantum system.

Furthermore, for completeness, we observe that the modification of the
kinetic term with the inclusion of long-range hopping is expected to
lead to multiple Majorana end modes both in spinless
\cite{DeGottardi2013a} and spinfull $p$-wave SC chains
\cite{Mercaldo2016}.
We also point out that for the case of intrinsic diluted pairing centers,
the model Hamiltonian in Eq. 1 would strictly apply only in the weak-coupling regime.
Therefore, it is challenging to investigate to what extent the overall scenario
will be modified by including finite NN interactions without any decoupling \cite{Kat15}.

Various realizations of the presented topological states may be possible.
One way is to design a mesoscopic array
of SC dots coupled to a metallic host in such a way that
nearby the dot an effective spin-triplet pairing is induced,
e.g. by employing spin-orbit and external magnetic field. Here, a local tuning of the
superconducting amplitude can be achieved by external perturbations, while the hopping connectivity among
the pairing centers might be tailored by the distance between the dots, or by suitably
varying the conducting channels linking the dots. Another
possible realization is in doped 1D spin-orbital quantum systems
where coupling between host and dopants could be converted into
effective pairing terms for the spin or the orbital channel.
One could then atomically design doped quantum
chains that map onto the Kitaev model \cite{Kitaev2001} with diluted
PCs. Then, the present study may stimulate new directions of
research on
the generation and manipulation of
Majorana states and their design for topological quantum computing
\cite{Iva01,Nay08}.

We kindly acknowledge support by Narodowe Centrum Nauki
(NCN) under Project No.~2012/04/A/ST3/00331.
W. B. acknowledges financial support by the European Union's Horizon 2020
research and innovation programme under the Marie
Sklodowska-Curie Grant Agreement No. 655515.
M. C. acknowledges support of the Future and Emerging Technologies (FET)
programme under FET-Open Grant No. 618083 (CNTQC).


\clearpage

\section{Supplemental Material}

In this Supplemental Material we present more details on the expansion
of ${\cal A}$ in powers of $\{y_i\}$, link between the cases of uniform
and alternating $\mu_{0}$ in the host and the generic form of
coefficient ${\cal A}$ in case of arbitrary hopping amplitudes $t_{p}$,
pairing amplitudes $\Delta_{p}$ and chemical potentials $\mu_{p}$ on
every bond and site of the unit cell. Furthermore, we provide a zoomed
view of the interference regions of the disordered phase diagram close
to the resonant condition $y_i\equiv y_{\rm imp}$.
Finally, a representative case for a fully disordered configuration is
shown to demonstrate the occurrence and the character of the Majorana
modes at the end sites of a chain of finite size.

\subsection{A. Expansion of ${\cal A}$}

Coefficients ${\cal B}$ and ${\cal C}$ (4) are rather easy to obtain
so we first focus on ${\cal A}$. ${\cal A}$ is a combination of
three determinants of tridiagonal matrices obtained from ${\bf u}_{k}$.
These determinants can be further Laplace-expanded to separate the
parameters of the impurities from the pure-host determinants. We set
a uniform host, i.e., $\mu_{p}=\mu_{0}$ and $t_{p}=t_{0}$ for all
host sites and bonds. The pure host determinant $D_{q}$ for $q$
host's atoms between two impurities has a form of,
\begin{equation}
D_{q}=\det\begin{pmatrix}\mu_{0} & t_{0} & 0 & 0\\
t_{0} & \ddots & \ddots & 0\\
0 & \ddots & \mu_{0} & t_{0}\\
0 & 0 & t_{0} & \mu_{0}
\end{pmatrix}.
\end{equation}
Now it is important to notice that such a tridiagonal matrix can be
diagonalized in terms of modes of a particle in a box with the
resulting eigenvalues
\begin{equation}
\lambda_{q,n}=\mu_{0}-2t_{0}\cos\left(n\frac{\pi}{q+1}\right),
\end{equation}
with $n=1,2,\dots,q$. The determinant is thus a product of these
eigenvalues but it can be rephrased in terms of a trigonometric
expression,
\begin{equation}
D_{q}=\frac{2\sin\left[(q+1)\eta_{0}\right]}{\sqrt{4-\mu_{0}^{2}/t_{0}^{2}}},
\label{eq:dethost}
\end{equation}
with
\begin{equation}
\eta_{0}=\arccos\frac{\mu_{0}}{2t_{0}}.\label{eq:eta0}
\end{equation}
It is worth to notice that in fact
$D_{q}\equiv U_{q}\left(\mu_{0}/2t_{0}\right)$, where $U_{n}(x)$ is
a Chebyshev polynomial of the second kind of rang $n$.
Note that $D_{q}$ is defined not only for $|\mu_{0}|<|2t_{0}|$ but also
for any other $\mu_{0}$ by an analytical continuation of the arccos
function. Similarly, the zeroth order term in $\{y_i\}$ appearing in
the expansion of ${\cal A}$ is also a polynomial in $(\mu_{0}/2t_{0})$,
i.e., the $L$-th Chebyshev polynomial of the first kind,
$\cos(L\eta_{0})\equiv T_{L}\left(\mu_{0}/2t_{0}\right)$.

\subsection{B. Alternating $\mu_{0}$ by a 'Wick rotation'}

\begin{figure*}[t!]
\includegraphics[clip,width=0.77\textwidth]{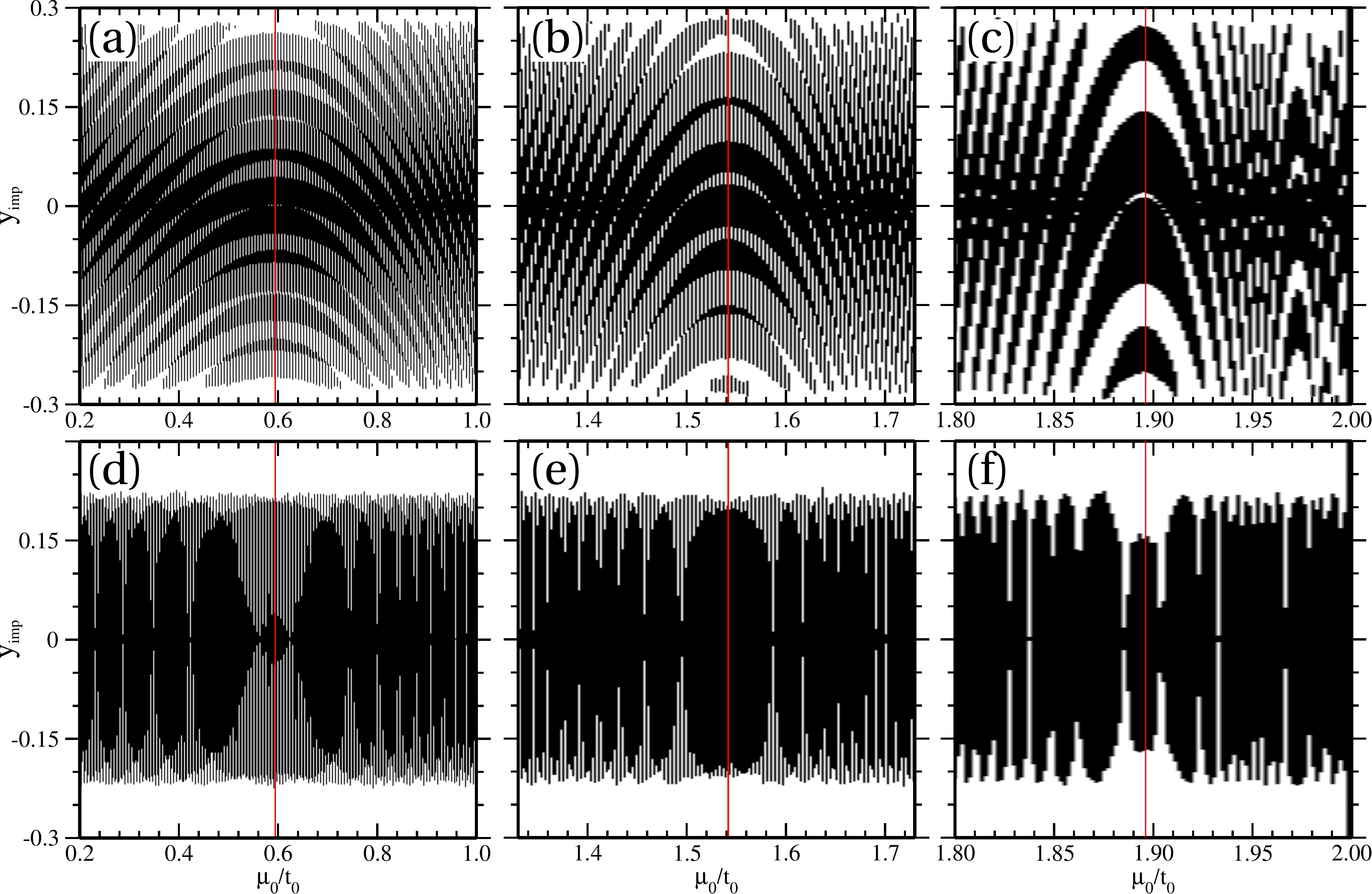}
\protect\caption{
Zoomed view of the topological (black) domains of Fig. 3(c)
(first row) and 3(d) (second row) of the main text close to specific
values of $\mu_0/t_0$ (marked with red lines).
The panels (a)-(c) refer to systems with a unit cell of $L\!=\!1000$
and $N\!=\!20$ randomly distributed impurities having the same
$y_i\equiv y_{\rm imp}$ for $\mu_0/t_0\simeq 0.6,1.54,1.9$,
respectively. The panels (d)-(f) refer to the same system size of
(a)-(c) but with the inclusion of a random sign of $\{y_i\}$ around
the same values of $\mu_0/t_0$.
\label{fig:zoom}}
\end{figure*}

Up to now we were working only in the uniform host limit where for all
$\mu_{p}=\mu_{0}$ for all host's sites. It is however surprisingly
easy to switch to the case where the chemical potential of the host is
alternating as $\mu_{p}=(-1)^{p}\mu_{0}$. Assume that $L$ is even,
which is necessary, and for simplicity that there is only one impurity
in the unit cell at site $p=L$. We get an anti-diagonal block
${\bf u}_k$ of ${\cal H}_k$ as,
\begin{equation}
{\bf u}_{k}=-\begin{pmatrix}
\mu_{1} & \tau_{L-1}^{-} & 0 & \cdots & e^{-ik}\tau_{L}^{+}\\
\tau_{L-1}^{+} & -\mu_{0} & \ddots & 0 & \vdots\\
0 & \ddots & \ddots & \tau_{2}^{-} & 0\\
\vdots & 0 & \tau_{2}^{+} & -\mu_{0} & \tau_{1}^{-}\\
e^{ik}\tau_{L}^{-} & \cdots & 0 & \tau_{1}^{+} & \mu_{0}
\end{pmatrix},
\end{equation}
with $\tau_p^{\pm}\equiv t_p\pm\Delta_p$.
Now we notice that the alternation in $\mu_{0}$ can be removed by
multiplying ${\bf u}_{k}$ from left and from the right by a diagonal
matrix ${\bf \sigma}$ having a diagonal of the form
$\left\{ 1,i,1,i,\dots,1,i\right\}$.
Denoting $\tilde{{\bf u}}_{k}={\bf \sigma}{\bf u}_{k}\sigma$ we get,
\begin{equation}
\tilde{{\bf u}}_{k}=-\begin{pmatrix}
\mu_{1} & i\tau_{L-1}^{-} & 0 & \cdots & ie^{-ik}\tau_{L}^{+}\\
i\tau_{L-1}^{+} & \mu_{0} & \ddots & 0 & \vdots\\
0 & \ddots & \ddots & i\tau_{2}^{-} & 0\\
\vdots & 0 & i\tau_{2}^{+} & \mu_{0} & i\tau_{1}^{-}\\
ie^{ik}\tau_{L}^{-} & \cdots & 0 & i\tau_{1}^{+} & \mu_{0}
\end{pmatrix}.
\end{equation}
This means that we are back in the uniform case because the determinant
of ${\bf u}_{k}$ is proportional to the one of $\tilde{{\bf u}}_{k}$!
More precisely, in the alternating case we are allowed to use the
formulas for ${\cal A}$, ${\cal B}$ and ${\cal C}$ which were derived
for the uniform case provided that we transform the parameters in
the following way:
\begin{eqnarray}
\mu_{0} & \to & \mu_{0}  \nonumber \\
t_{0} & \to & -it_{0}
\label{eq:wick_host}
\end{eqnarray}
for the host's parameters and
\begin{eqnarray}
\mu_{i} & \to & (-1)^{p_{i}}\mu_{i}\nonumber \\
t_{i} & \to & -it_{i}\nonumber \\
\Delta_{i} & \to & -i\Delta_{i}\label{eq:wick_imp}
\end{eqnarray}
for the impurities' parameters, where $p_{i}$ is the position of
the impurity $i$. Note that apart from the chemical potentials of
the impurities this is indeed a Wick rotation of ${\cal A}$, ${\cal B}$
and ${\cal C}$ in terms of hopping and pairing amplitudes. It is
straightforward to check that all these quantities remain real after
such transformation.

Note that the main difference between the uniform and alternating case
is as follows: For the uniform case the ${\cal A}$ coefficient in the
limit of $|\mu_{0}|<|2t_{0}|$ is a sum of many sine and cosine
functions that depend on angle $\eta_{0}$ so there are many
interference terms expected and the profile of ${\cal A}$ is indeed
very complicated. But in the opposite limit of $|\mu_{0}|>|2t_{0}|$ the
trigonometric function become hyperbolic and consequently ${\cal A}$
will be roughly exponential in $\eta_{0}$ following the highest
exponent $L\eta_{0}$. On the other hand, in the alternating regime,
${\cal A}$ will consist of hyperbolic functions for any values of
$\mu_0$ so its profile will be much simpler. Since having a topological
phase means having small ${\cal A}$, it will be much more difficult in
the alternating regime than in the uniform one.

\subsection{C. Coefficient ${\cal A}$ for arbitrary $\{t_{p},\Delta_{p},\mu_{p}\}$
on every bond and site}

Setting the hyperbolic parametrization of the hopping and pairing
amplitudes,
\begin{equation}
t_{p}=r_{p}\cosh\phi_{p}, \hskip .7cm \Delta_{p}=r_{p}\sinh\phi_{p},
\end{equation}
for all sites and bonds in the unit cell one can show that the
coefficient ${\cal A}$ can be expressed in terms of chemical potential
and hyperbolic radii only as,
\begin{equation}
{\cal A}=\frac{1}{2r_{1}r_{2}\cdots r_{L}}\begin{vmatrix}-\mu_{1} & r_{1} & 0 & 0 & r_{L}\\
r_{1} & -\mu_{2} & r_{2} & 0 & 0\\
0 & r_{2} & -\mu_{3} & \ddots & 0\\
0 & 0 & \ddots & \ddots & r_{L-1}\\
r_{L} & 0 & 0 & r_{L-1} & -\mu_{L}
\end{vmatrix}+\left(-1\right){}^{L}.
\end{equation}
Note that this expression has an explicit translation symmetry within
the unit cell. The remaining ${\cal B}$ and ${\cal C}$ coefficients
have an already the known form of Eqs. (4).
Note that the hyperbolic parametrization not only covers the area of
$t_{p}\ge0$ and $|\Delta_{p}|<t_{p}$ but can be extended to the whole
$t_{p}-\Delta_{p}$ plane (excluding the singular lines of
$|\Delta_{p}|=|t_{p}|$) by analytical continuation of the inverse
expressions, $r_{p}=\sqrt{t_{p}^{2}-\Delta_{p}^{2}}$ and
$\phi_p={\rm arctanh}\left(\Delta_p/t_p\right)$, in the complex plane.

\subsection{D. Additional features for a random distribution of impurities:
zoomed view of the topological phase diagram and Majorana end modes}

\begin{figure}[t!]
\includegraphics[clip,width=1\columnwidth]{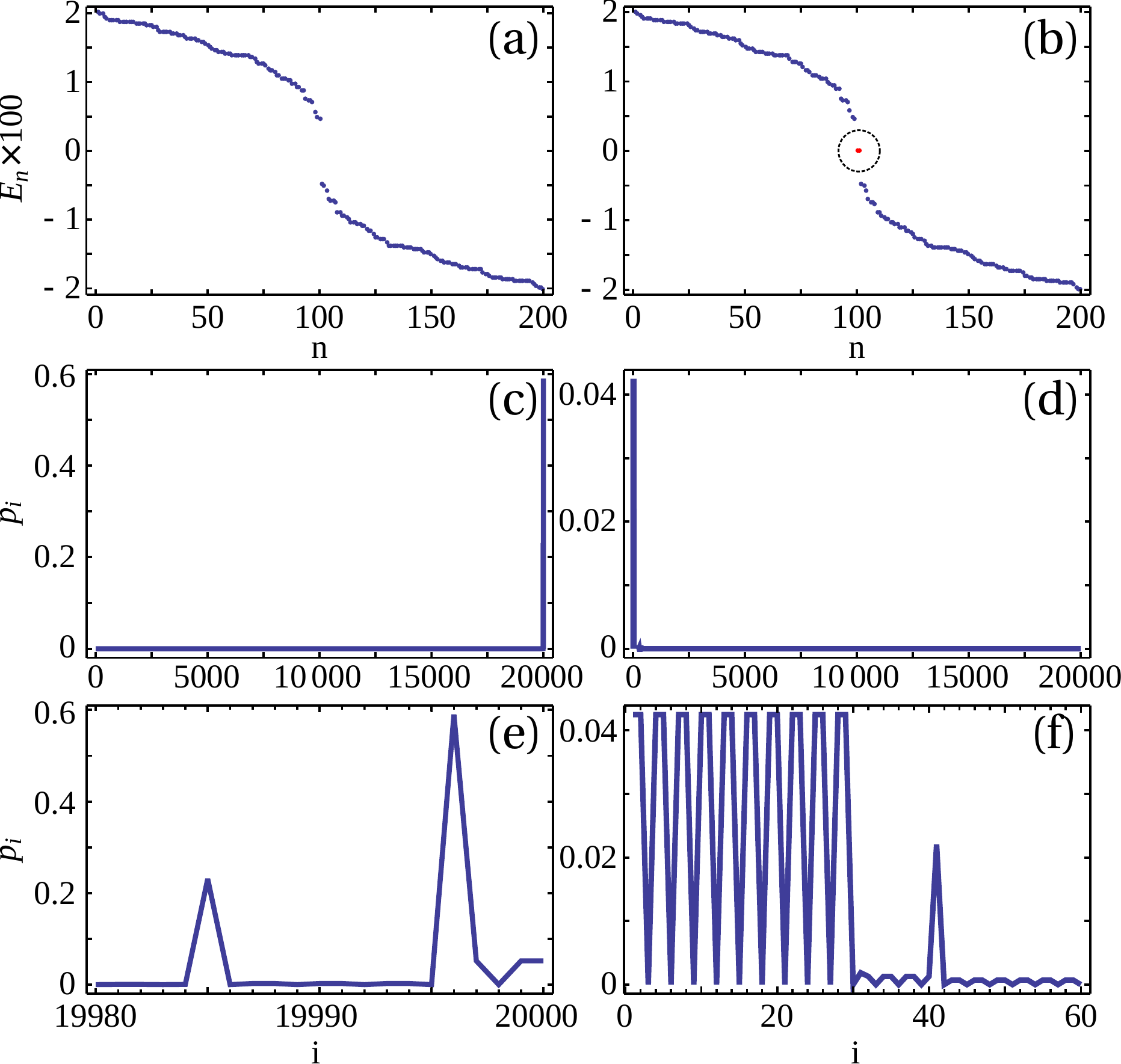}
\protect\caption{
Energy spectra and edge states for a fully disordered 1D chain of size
$L\!=\!20000$ and $N\!=\!200$ impurities assuming they are randomly
distributed in space and the parameters correspond to $\{y_i\}=0$.
(a)-(b) Spectra around zero energy for a closed system:
(a) and the same spectrum for an open system and
(b) exhibiting two zero energy states in the gap.
Panel (c) indicates the occupation probability for the zero energy
Majorana wave function versus the position $i$ at the right chain edge,
with (e) being a zoomed view. (d) shows the occupation probability of
the Majorana zero energy state projected at each position $i$ of the
left chain edge. A zoomed view is given in (f).
\label{fig:estates}}
\end{figure}

In Fig. \ref{fig:zoom} we show the details of Figs. 3(c) and 3(d) as
presented in the main text. A zoomed view of the topological domains
around three selected values of $\mu_0/t_0$, i.e.,
$\mu_0/t_0\simeq 0.6,1.54,1.9$ is provided to identify better
the observed self-similar features. In case of random positions of
impurities and fixed $y_i\equiv y_{\rm imp}$ we see a
parabolic pattern repeating itself
which becomes increasingly narrow as we approach
$\mu_0=2t_0$. This narrowing is related with the functional character
of $\eta_0=\arccos(\mu_0/2 t_0)$, i.e., it has a singular derivative
near $\mu_0=2t_0$.
In case of random positions of the impurities and random sign of
$\{y_i\}$ we do not observe such self-similarity but still distinct
features are visible at $\mu_0/t_0\simeq 0.6,1.54,1.9$.
We note that the horizontal resolution of the plots effectively
decreases as we increase $\mu_0/t_0$ because we have an increasingly
narrow window of $\mu_0/t_0$.

In Fig. \ref{fig:estates} we report a representative case where the
Majorana bound states appear at the end of the chain in a finite system
when the condition for the topological invariant to be non-trivial is
fulfilled. Indeed, we determine the energy spectra and the edge states
for a system of $L\!=\!20000$ with $N\!=\!200$ randomly distributed
impurities having all $\{y_i\}$ at the resonant condition $y_i=0$.
This regime does not however imply that the impurities are identical,
i.e., in terms of $\{t_i,\Delta_i,\mu_i\}$.

Using the hyperbolic parametrization, $t_i=r_i\cosh \phi_i$ and
$\Delta_i=r_i\sinh \phi_i$, we can achieve the resonance condition
$y_i=0$ by setting the host's parameters at $t_0=\mu_0=1$ and the
impurity's chemical potential $\mu_i$ as $\mu_i=r_i^2$, see Eq. (3) in
the main text. Thus, we are left with two free parameters, $r_i$ and
$\phi_i$ at every impurity site. Then, one can randomly tune these
amplitudes within the interval $[0,1]$. Here, we avoid negative angles
for $\phi_i$ as they would lead to a vanishing $\sum_i \phi_i$ and, in
turn, of the coefficient $\cal{C}$ (4), thus violating the necessary
condition for having a fully gapped bulk spectrum.

In Figs. \ref{fig:estates}(a) and \ref{fig:estates}(b) we present the
spectra around zero energy for a closed and open system, respectively.
We note that for an open system there are two zero-energy states
appearing in the energy gap. These are Majorana end modes that arise as
a consequence of the bulk-boundary correspondence in a topologically
non-trivial configuration.
In Figs. \ref{fig:estates}(c) and \ref{fig:estates}(d) the spatial
occupation probabilities in the two zero energy states is explicitly
shown in order to confirm their degree of localization on the
right/left edges of the 1D chain.
Finally, in Figs. \ref{fig:estates}(e) and \ref{fig:estates}(f) we
provide a zoomed view
of the spatial distribution function just close to
ends of the chain at the interface with the vacuum. As one can notice,
the occupation probability for the two states is not symmetric due to
the inversion symmetry breaking induced by the presence of a disordered
pattern of pairing centers. Indeed, the occupation pattern reveals that
Majorana wave function has an internal structure and involves only few
sites for the left edge while it extends over more sites for the right
edge boundary.

\end{document}